\documentclass[sn-mathphys,Numbered]{sn-jnl}

\usepackage{graphicx}%
\usepackage{multirow}%
\usepackage{amsmath,amssymb,amsfonts}%
\usepackage{amsthm}
\usepackage{mathrsfs}%
\usepackage[title]{appendix}%
\usepackage{xcolor}%
\usepackage{textcomp}%
\usepackage[legacycolonsymbols]{mathtools}
\usepackage{manyfoot}%
\usepackage{booktabs}%
\usepackage{algorithm}%
\usepackage{algorithmicx}%
\usepackage{algpseudocode}%
\usepackage{listings}%
\usepackage{ulem}%

\begin{document}

\title[Article Title]{A Design Method of an Ultra-Wideband and Easy-to-Array Magic-T: A 6-14 GHz Scaled Model for a mm/submm Camera}


\author*[1,2]{\fnm{Shuhei} \sur{Inoue}}\email{sinoue@ioa.s.u-tokyo.ac.jp}
\author[1,2]{\fnm{Kah Wuy} \sur{Chin}}
\author[1,2]{\fnm{Shinsuke} \sur{Uno}}
\author[1]{\fnm{Kotaro} \sur{Kohno}}
\author[3]{\fnm{Yuka} \sur{Niwa}}
\author[2,4]{\fnm{Toyo} \sur{Naganuma}}
\author[2,4]{\fnm{Ryosuke} \sur{Yamamura}}
\author[2,5]{\fnm{Kazuki} \sur{Watanabe}}
\author[6]{\fnm{Tatsuya} \sur{Takekoshi}}
\author[2,5]{\fnm{Tai} \sur{Oshima}}


\affil*[1]{\orgdiv{Institute of Astronomy, Graduate School of Science}, \orgname{The University of Tokyo}, 
\orgaddress{\street{2-21-1 Osawa}, \city{Mitaka}, \postcode{181-0015}, \state{Tokyo}, \country{Japan}}}

\affil[2]{\orgdiv{Advanced Technology Center}, \orgname{National Astronomical Observatory of Japan}, \orgaddress{\street{2-21-1 Osawa}, \city{Mitaka}, \postcode{181-8588}, \state{Tokyo}, \country{Japan}}}

\affil[3]{\orgname{Tokyo Institute of Technology}, \orgaddress{\street{2-12-1 Ookayama}, \city{Meguro-ku}, \postcode{152-8551}, \state{Tokyo}, \country{Japan}}}

\affil[4]{\orgdiv{Graduate School of Informatics and Engineering}, \orgname{The University of Electro-Communications}, \orgaddress{\city{Cho-fu}, \postcode{152-8551}, \state{Tokyo}, \country{Japan}}}

\affil[5]{\orgdiv{Graduate Institute for Advanced Studies}, \orgname{SOKENDAI}, \orgaddress{\street{2-21-1 Osawa}, \city{Mitaka}, \postcode{181-8588}, \state{Tokyo}, \country{Japan}}}

\affil[6]{\orgname{Kitami Institute of Technology}, \orgaddress{\street{165 Koen-cho}, \city{Kitami}, \postcode{090-8507}, \state{Hokkaido}, \country{Japan}}}


\abstract{We established a design method for a Magic-T with a single-layer dielectric/metal structure suitable for both wideband and multi-element applications for millimeter and submillimeter wave imaging observations.
The design method was applied to a Magic-T with a coupled-line, stubs, and single-stage impedance transformers in a frequency-scaled model (6--14 GHz) that is relatively easy to demonstrate through manufacturing and evaluation.
The major problem is that using the conventional perfect matching condition for a coupled-line alone produces an impractically large width coplanar coupled-line (CPCL) to satisfy the desired bandwidth ratio. In our study, by removing this constraint and optimizing impedances utilizing a circuit simulator with high computation speed, we found a solution with a $\sim$ 180 $\rm \mu m$ wide CPCL, which is approximately an order of magnitude smaller than the conventional analytical solution.
Furthermore, considering the effect of transition discontinuities in the transmission lines, we optimized the line length and obtained a design solution with return loss $<$ -20 dB, amplitude imbalance $<$ 0.1 dB, and phase imbalance $<$ 0.5$^{\circ}$ from 6.1 GHz to 14.1 GHz.}

\keywords{Magic-T, Ratrace, Wideband, Single-layer, Coplanar, Coupled-line}


\maketitle
\section{Introduction}\label{sec1}
In recent years, in order to perform wideband millimeter and submillimeter continuum wave imaging (typically cosmic microwave background observations), superconducting detector arrays have been developed in which planar antenna and readout elements are integrated on a dielectric substrate and 
arranged on a focal plane (see review \cite{2018Hubmayr}).

A Magic-T combines two signals with a phase difference of 0$^{\circ}$ or 180$^{\circ}$ at the sum ($\Sigma$) port or difference ($\Delta$) port, respectively, and often used for dual polarization observation with an ortho-mode transducer (OMT), which separates each polarization with two pairs of probes. For example, it combines two signals of the fundamental mode (TE11) at $\Delta$ port, which are in opposite phase and sent through the optics and horn antenna to an OMT on the detector substrate. The Magic-T also terminates undesirable higher-order even modes at the $\Sigma$ port, which serves to shape the beam \cite{2014Datta}.

In previous studies, an ultra-wideband over BWR $\geq 2.3$ (BWR:= bandwidth ratio, the highest/lowest frequency ratio in a band) have been realized using a multi-layer structure or a crossover to a narrow stripline requiring high accuracy alignment \cite{2012Gruszczynski, 1999Chang}. 
On the other hand, for detector array applications, it is desirable for the Magic-T to be manufactured as a single layer dielectric/metal (hereafter referred to as \textit{single-layer}), which requires fewer manufacturing processes and is easier to fabricate.

In this paper, we present a design method for a Magic-T that is both ultra-wideband and single-layer, without using a crossover to a stripline.

\section{Wideband and single-layer Magic-T}\label{sec2}
A well-known basic structure of Magic-T is a ratrace coupler consisting of three $\lambda$/4 ($\lambda$:= wavelength) and a 3$\lambda$/4 transmission lines \cite{1947Tyrrell}. The bandwidth is narrow because the phase inversion depends on the difference in line length. Then, a wideband Magic-T was proposed to replace the 3$\lambda$/4 line with a shorted coupled-line, which enables phase inversion by capacitive coupling \cite{1968March}. Moreover, further broadbanding is demonstrated using stubs that compensate for the asymmetric short stubs that the coupled-line includes, and multi-stage impedance transformers (ITFs) that suppress impedance mismatch \cite{2012Gruszczynski}. Theoretically, zero amplitude and phase imbalance ($\delta A$ and $\delta \theta$; deviation from appropriate amplitude and phase difference at $\Sigma$ or $\Delta$ port), perfect isolation (IL; signal leakage between two input ports.  "perfect" means -$\infty$ dB), return loss (RL; signal loss due to reflection at the input port) $\leq$ -20 dB are obtained over BWR $\geq 2.3$.

The above analytical solution uniquely determines the impedance solution by imposing the perfect matching condition on the coupled-line alone \cite{2012Gruszczynski}. However, when using a coplanar coupled-line (CPCL), which is a single-layer and does not require a crossover to a stripline, it is difficult to realize the analytical solution because it requires a wide full width ($:= g + 2w + 2s$, see section \ref{sec3}, figure \ref{fig1}(a,b)) close to the ratrace scale. However, the perfect matching condition can be removed \cite{1997Walker}, and even in that case perfect isolation and zero imbalance are preserved.

With these in mind, we propose a method to obtain a solution that narrows the full width of a CPCL by removing the perfect matching condition expanding the range of impedance solutions, and exhaustively searching by utilizing a fast circuit simulator.

\section{Requirements and basic structure}\label{sec3}

\begin{figure}[ht]%
\centering
\vspace{-9.0mm}
\includegraphics[width=1.0\textwidth]{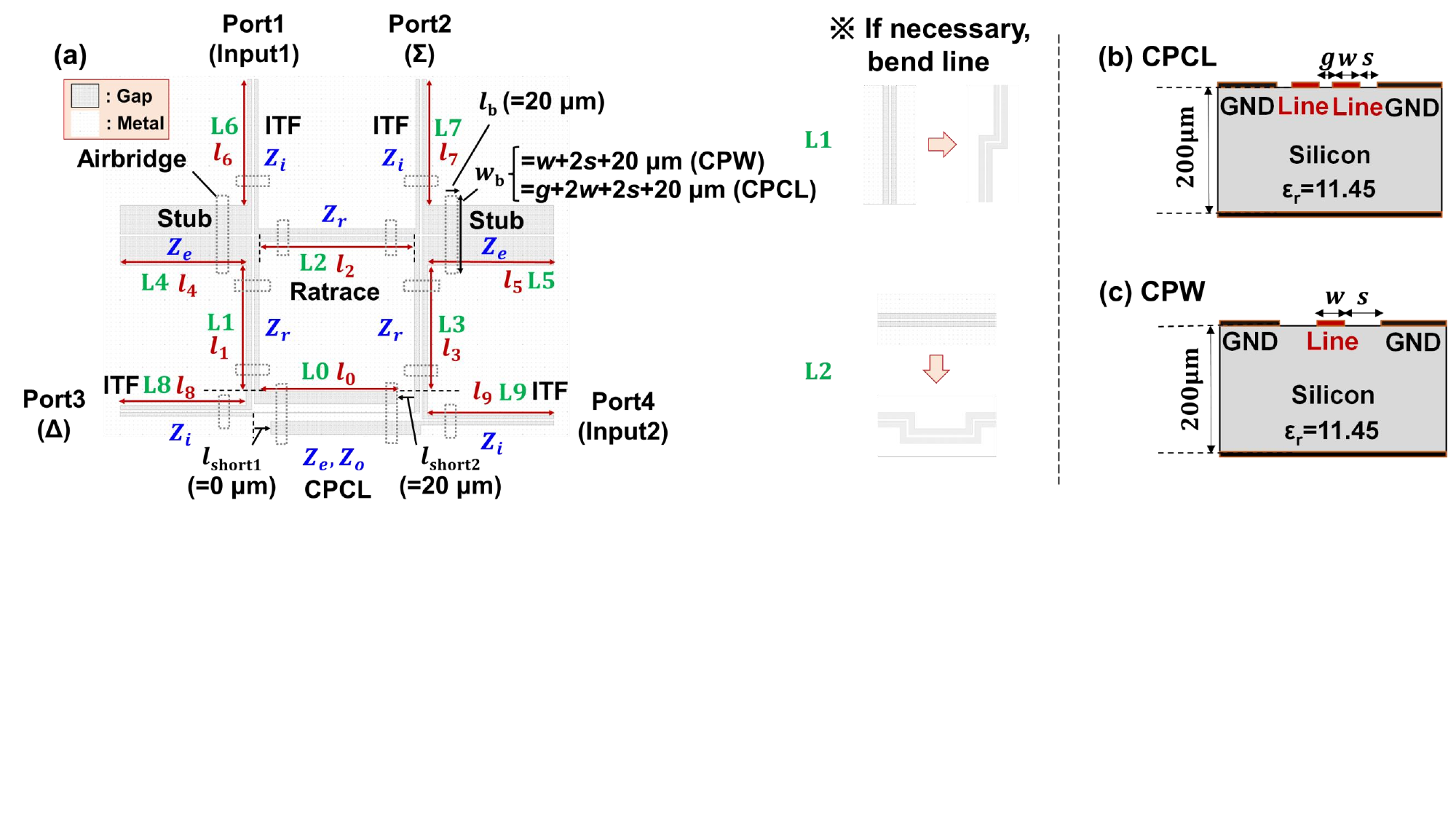}
\vspace{-32.0mm}
\caption{(a): The basic structure of the Magic-T used in this study (top view; not to scale) and definitions of impedance (blue),
number of transmission lines (green), and line lengths (brown). The height of the airbridge $h_{\rm b}$ is fixed at 10 $\mu$m. Note that L1 and L2 are bent according to the optimization results. (b,c): Definition of line width and gap space of the CPCL and CPW (cross-sectional view). $\epsilon_{\rm r}$ means relative permittivity of silicon.} \label{fig1}
\end{figure}

Here we summarize the design conditions of our Magic-T. The bandwidth ratio criterion was given to be BWR = 2.3. This value corresponds to the limited instantaneous bandwidth of horn-coupled OMTs \cite{2017abitbol}. For return loss and phase imbalance, the values obtained in \cite{2012Gruszczynski} are used as the standard, and amplitude imbalance is set to be equivalent to the amplitude deviation due to phase imbalance. From the above, it was determined that the requirements for the Magic-T are RL $\leq$ -20 dB, $\delta \theta$ $\leq$ 1.5$^\circ$, and $\delta A$ $\leq$ 0.2 dB be satisfied over BWR $\geq 2.3$.
In this study, to verify the design method through relatively simple manufacturing and performance evaluation in the future, the center frequency was set to 10 GHz and the minimum line width of the transmission line was set to 3 $\mu$m, which can be fabricated reliably using stepper lithography. Furthermore, the input impedance of the Magic-T is 100 $\Omega$ to match the impedance of  CPW feed lines connected to the OMT probe via impedance transformation.

In order to meet the required BWR $\geq 2.3$, the basic structure of our Magic-T adopted the circuit model proposed in \cite{2012Gruszczynski}, which includes an coupled-line, stubs, and 1-stage ITFs. A schematic diagram of the structure is shown in Fig. \ref{fig1}(a). Here, the even and odd mode impedances of the coupled-line are $Z_{\rm e}$ and $Z_{\rm o}$, respectively, the impedance of the ratrace section is $Z_{\rm r}$, and the impedance of the ITF is $Z_{\rm i}$. Due to the matching condition of the ratrace section, $Z_{\rm r}=2Z_{\rm e}Z_{\rm o}/(Z_{\rm e}-Z_{\rm o})$ is fixed. In addition, the impedance of the stub is fixed to $Z_{\rm e}$ so that theoretically zero imbalance is achieved. 
The line numbers are defined as shown in Fig. \ref{fig1}(a; green characters). As transmission lines, a CPCL and coplanar waveguides (CPWs) are used and the line width and the gap space are defined as shown in Fig. \ref{fig1}(b,c). Since the full width increases as $g$ of the CPCL increases, we fix $g = 3$ $\mu$m in this study. The substrate dielectric is assumed to be a silicon wafer (permittivity $= 11.45$ \cite{2006Krupka}, height $= 200$ $\mu$m), and dielectric losses are assumed to be negligible. Niobium was assumed as a superconductor, and the kinetic inductance is about one order of magnitude smaller than the CPW geometric inductance (calculated from supplementary of \cite{2022Hahnle}). To save computation time, we neglected the sheet impedance and film thickness.

\section{Search for coplanar coupled-line solutions}\label{sec4}

\begin{figure}[ht]%
\centering
\vspace{-2.0mm}\includegraphics[width=0.4\textwidth]{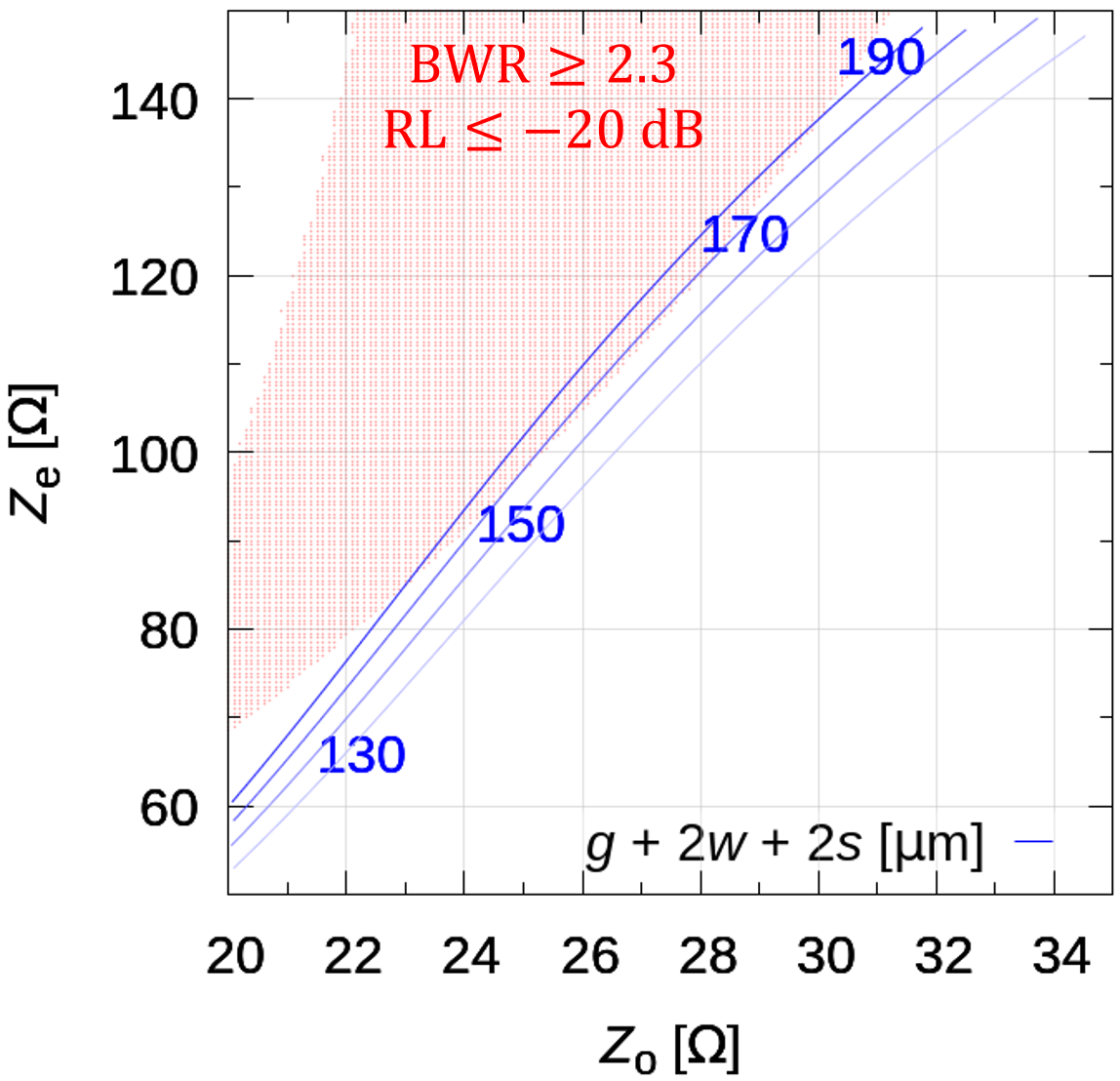}
\caption{Full width of a CPCL and bandwidth ratio of the Magic-T versus impedance $Z_{\rm e}$ and $Z_{\rm o}$ of the coupled-line. (Blue line): Full width calculated from the analytical solution for $g=3$ $\mu$m \cite{1985Hanna}. (Red region): Region of impedance above the bandwidth ratio of 2.3 calculated by the equivalent circuit model of QUCS. Contours with full width $\geq$ 170 $\rm \mu m$ overlap the region where BWR $\geq 2.3$.}\label{fig2}
\end{figure}

The full width range of the CPCL that can achieve BWR $\geq 2.3$ was estimated by removing the perfect matching condition ($Z_{\rm e}Z_{\rm o}=Z_{\rm r}^2$) for the coupled-line alone. Transmission line models of Quite Universal Circuit Simulator (QUCS) were used in calculations of the Magic-T performance. The relationship between impedance and full width of a CPCL was calculated using analytical solution \cite{1985Hanna}. The impedance search range in the calculation of bandwidth ratio was set to $Z_{\rm e} = $ 50-150-0.5 $\Omega$, $Z_{\rm o}=$ 15-45-0.1 $\Omega$ and $pZ_{\rm i} \equiv Z_{\rm i}/Z_{\rm r} =$ 0.60-1.30-0.002 (min.-max.-step). As for the bandwidth ratio, the widest one among the $pZ_{\rm i}$ search ranges was selected. Note that $pZ_{\rm i}$ was taken to be close to 1, and it was confirmed that optimal solution was not obtained at the edge of the range of $pZ_{\rm i}$. Here, the effective even mode and odd mode permittivity of the coupled-line ($\epsilon_{\rm re}$ and $\epsilon_{\rm ro}$) were assumed to be both 1, and the electrical length was $\lambda$/4, which is an ideal case. Fig. \ref{fig2} shows the relationship between the full width and the region where BWR $\geq 2.3$ for each $Z_{\rm e}$ and $Z_{\rm o}$ in the above search range.

From Fig. \ref{fig2}, we confirmed that the lower limit of the CPCL full width is about 170 $\mu$m. This is about one order of magnitude lower than the analytical solutions $Z_{\rm e}=163.2$ $\Omega$ and $Z_{\rm o}=28.0$ $\Omega$ given by \cite{2012Gruszczynski}. Based on the results, we will attempt to optimize the impedance and line length of the Magic-T in the next section.

\section{Optimization}\label{sec5}
In this section, we first determine the cross-sectional CPCL and CPWs geometries by optimizing impedances based on bandwidth and full width criteria. Then we determine $l_1$, $l_2$, $l_3$, $l_4$, $l_6$, $l_9$ by minimizing the imbalance. Lastly we determine $l_5$, $l_7$ and $l_8$ to minimize both return loss and insertion loss.

\subsection{Impedance optimization}\label{subsec51}
The impedances of the Magic-T were optimized using QUCS to determine all line widths. Electrical lengths of CPWs were fixed to $\lambda$/4. For CPCL, the even and odd mode permittivities are slightly different. Therefore, the physical length was adjusted so that for the ideal model with the $\lambda$/4 electrical lengths for both modes, the sum of squared residuals of the S-matrix is minimized. Under these conditions, in order to prevent not only a CPCL but also stubs from becoming wider, we searched for a pair of impedances that minimizes the full width of a CPCL and stubs that satisfy the requirements. Referring also to the results obtained in section \ref{sec4}, independent variables and their ranges were $Z_{\rm e}=$80-110-0.1 $\Omega$, $Z_{\rm o}=$21-28-0.1 $\Omega$, $pZ_{\rm i}=$0.8-1.2-0.005 (min.-max.-step). Note that the return loss criterion here is set to about -20.4 dB (= 0.095 mag) to provide margin for later optimization.

The optimized impedance and line widths are shown in table \ref{tab1}. The obtained optimal CPCL and stub linewidths are $(g,w,s)=(3.0, 41.9, 45.8)$ $\mu$m and $(w,s)=(3.0, 16.9)$ $\mu$m, respectively. Hence, the full width of the CPCL was less than 180 $\mu$m. Note that the line widths $w$ and $s$ of the CPCL were adjusted within +1.0 $\mu$m to match the impedance calculated by an electromagnetic simulator (Sonnet), respectively. Furthermore, it was confirmed that the optimal solution was not obtained at the edge of the impedance range. From Fig. \ref{fig3}(a), the solution satisfied the requirements, and the power leakage to $\Sigma$ port in odd mode input and to $\Delta$ port in even mode are $\leq$ -40 dB.

\subsection{Line length optimization}\label{subsec52}
\begin{figure}[ht]%
\centering
\vspace{-12.0mm}
\includegraphics[width=0.7\textwidth]{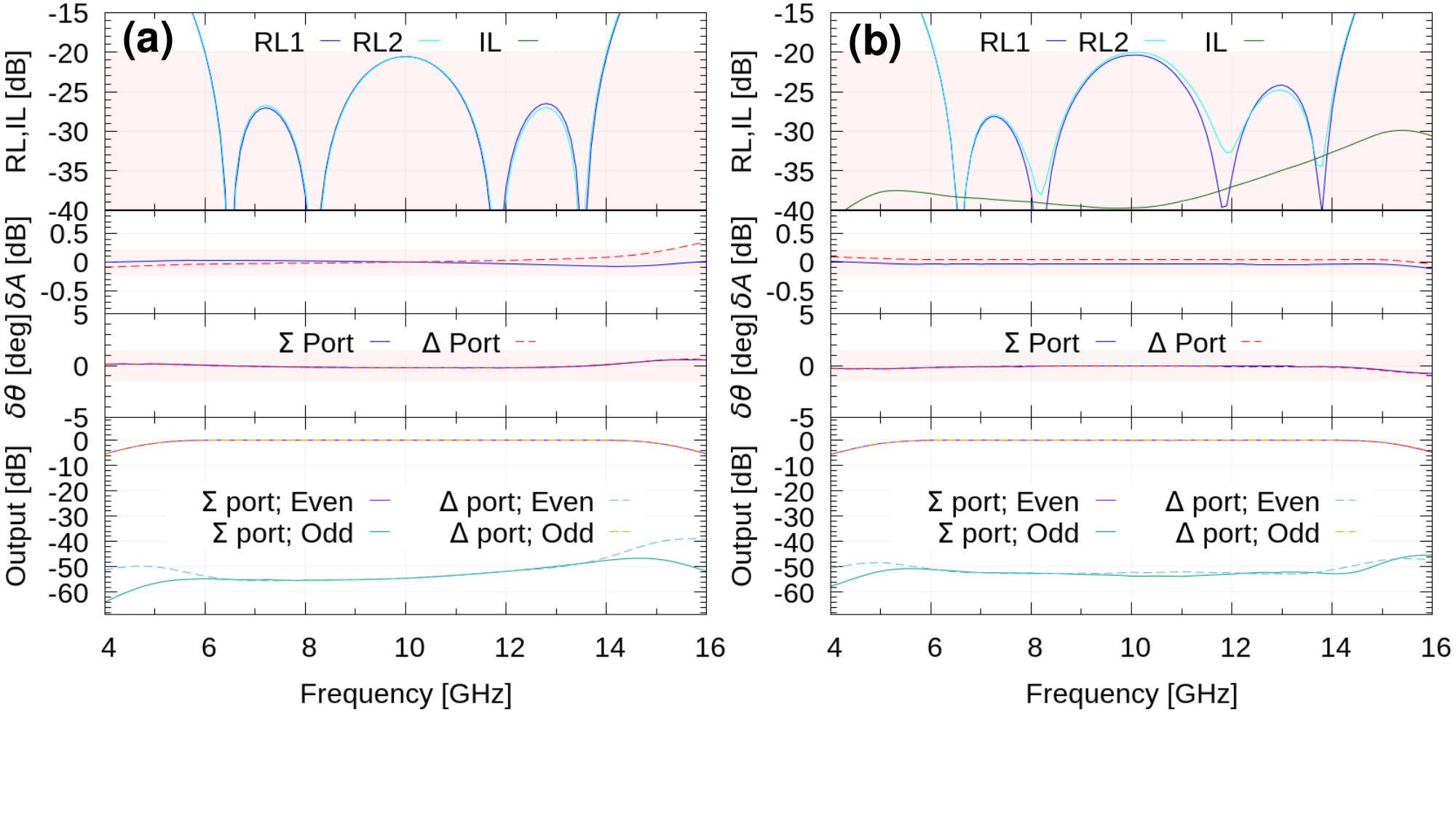}
\vspace{-7.0mm}
\caption{(a): Calculated performance of the Magic-T after impedance optimization with idealized transmission lines in QUCS. The effective permittivity of the coupled-line was calculated by equations from \cite{1985Hanna}. (b): Calculated performance after optimizing the transmission line length considering the transition. The S-matrices of all transmission lines and transitions are computed in Sonnet and cascaded in QUCS. (top): return loss (RL) in input1,2 and isolation (IL). (second): amplitude imbalance ($\delta A$), (third): phase imbalance ($\delta$$\theta$). (bottom): transmitted power in even and odd modes at the $\Sigma$ and $\Delta$ ports. Thin red boxes indicate the range where the required performance is satisfied. Note that the IL in (a) is not shown since it is less than -40 dB at all frequencies in the axis.}\label{fig3}
\end{figure}

As a preparation for the optimization of the line length ($l_0$--$l_9$), an initial estimate of the line length is determined by compensating for the discontinuity at the connection of lines (transition). S-matrices of the transitions between cross junctions in the ratrace section, between stubs and the GNDs, and between ports and the ITFs were calculated by Sonnet, respectively. The dimensions of the 100 $\Omega$ CPW extending from the port were ($w$,$s$) = (3, 22.7) $\mu$m.
In addition, the line lengths of the equivalent circuit model of each of the transitions were fitted so that the residual sum of squares of the S-matrices between the transition and the equivalent circuit model is minimized, and the initial values of the line lengths of the Magic-T were set to the values compensated for the deviations.
Then, the line length was exhaustively searched as follows. In the first step, $l_1$, $l_2$, $l_3$, $l_4$, $l_6$ and $l_9$, which affect the imbalance at the output, were determined to be the values that minimize the imbalance.
Next, $l_5$, $l_7$ and $l_8$, which have little effect on the imbalance at the output, were determined to minimize the in-band average 
of return loss and isolation power. Note that $l_0$ was fixed in the same way as in section \ref{subsec51}. The search range of line length was 
0.93-1.07-0.01 times (min.-max.-step) the initial value, and it was confirmed that the optimal solution was not obtained at the edge of the range. Finally, the idealized transmission lines in QUCS were replaced with CPWs and a CPCL S-matrices computed by Sonnet. The optimized line lengths are listed in Table.\ref{tab1}, and the performance is shown in Fig. \ref{fig3}(b).

\begin{table}[ht]
\caption{Optimized parameters in section \ref{sec5}. In section \ref{subsec51}, the impedance $Z$ and the dimensions $w$, $s$, and $g$ of each transmission line were determined, and in section \ref{subsec52}, the physical line lengths $l_p$ were determined. Note that the lengths are rounded to the nearest 10 µm, which is less than the step in the optimization.}\label{tab1}
\begin{tabular}{@{}rrrrrrrrrrrr@{}}
\toprule
Param. & L0 & L1 & L2 & L3 & L4 & L5 & L6 & L7 & L8 & L9\\
\midrule \midrule
$Z$ [$\rm \Omega$] & ($Z_{\rm e}, Z_{\rm o}$)=\\ & (93.5, 24.3) & 65.7 & 65.7 & 65.7 & 93.5 & 93.5 & 65.0 & 65.0 & 65.0 & 65.0 \\
 $w$ [$\rm \mu m$] & 41.9 & 3.0 & 3.0 & 3.0 & 3.0 & 3.0 & 3.0 & 3.0 & 3.0 & 3.0 \\
 $s$ [$\rm \mu m$] & 45.8 & 4.3 & 4.3 & 4.3 & 16.9 & 16.9 & 4.2 & 4.2 & 4.2 & 4.2\\
$g$ [$\rm \mu m$] &  3.0 & - & - & - & - & - & - & - & - & -\\
\midrule
$l_p$ [$\rm \mu m$] & 2900 & 2830 & 2950 & 2820 & 3020 & 3020 & 2980 & 2920 & 2950 & 2940 \\
\botrule
\end{tabular}
\footnotetext{}
\end{table}

Fig. \ref{fig3}(b) shows that even considering the transition discontinuity, RL $<$ -20 dB, IL $<$ -30 dB, $\delta$A $<$ 0.1 dB, and $\delta \theta$ $<$ 0.5$^\circ$ from 6.1 GHz to 14.1 GHz (BWR = 2.3), which satisfy the required performance. Moreover, the power leakage to the $\Sigma$ port in the odd mode and to the $\Delta$ port in the even mode can be negligible ($\sim$ -50 dB).

\section{Conclusion}\label{sec6}
We established a design method of a dielectric/metal single-layer Magic-T suitable for wideband and detector arraying, and applied this method to a scaled model Magic-T ($\sim$ 10 GHz).
The conditions of conventional analytical solutions were partially removed, and fast circuit simulations
were utilized to search for a solution with a narrower CPCL full width. Furthermore, we showed that a design solution that satisfies the requirements can be obtained by optimizing the impedance and line length of the Magic-T, taking into account the discontinuity of transitions.

In this study, the sheet impedance  was neglected to save the computation time. However, even if it is considered, almost the same method can be used by calculating the CPCL impedance in advance from an electromagnetic simulator.

Our next step is to fabricate and measure the design solution. In the future, the design method will be applied to millimeter and submillimeter waves. A potential challenge is that the shorter length of the ratrace requires a narrower CPCL full width (i.e., narrower center gap $g$). Due to manufacturing issues, it may be difficult to scale $g = 3$ $\mu$m in the 10 GHz model directly to millimeter or submillimeter waves. In this case, a tradeoff must be considered between a wider $g$ (i.e., easier fabrication) and larger transition discontinuities due to a wider CPCL full width. However, the performance degradation could be minimized using the established method, which can minimize the CPCL full width and compensate for the transition discontinuities.

\bmhead{Acknowledgments}
We would like to thank the anonymous referees for
their careful review and invaluable comments. We would like to express our deepest gratitude to T. Kobayashi, T. Taino, and S. Mima for their great cooperation from the viewpoint of fabrication. S.I. and S.U. are supported by FoPM, WINGS Program, the University of Tokyo. T.T. is supported by MEXT Leading Initiative for Excellent Young Researchers Grant Number JPMXS0320200188. This work was supported by JSPS KAKENHI Grant Numbers JP23H00121, JP23H01183, JP23H01209, JP21J20742.

\bibliography{sn-article.bib}
\end{document}